\documentclass[prl,twocolumn,showpacs,superscriptaddress,amsmath,amssymb,nofootnoteinbib]{revtex4}

\usepackage{graphicx}
\usepackage{dcolumn}
\usepackage{bm}

\begin{document}

\title{High electric field development for the SNS nEDM Experiment}

\author{T.~M.~Ito}
\email{ito@lanl.gov}
\affiliation{Los Alamos National Laboratory, Los Alamos, New Mexico 87545, USA}

\author{D.~H.~Beck}%
\affiliation{Loomis Laboratory of Physics, University of Illinois,
  Urbana, Illinois 61801, USA}%

\author{S.~M.~Clayton}
\affiliation{Los Alamos National Laboratory, Los Alamos, New Mexico 87545, USA}

\author{C.~Crawford}
\affiliation{Department of Physics and Astronomy, University of
  Kentucky, Lexington, Kentucky 40506, USA}

\author{S.~A.~Currie}
\affiliation{Los Alamos National Laboratory, Los Alamos, New Mexico 87545, USA}

\author{W.~C.~Griffith}
\affiliation{Los Alamos National Laboratory, Los Alamos, New Mexico 87545, USA}

\author{J.~C.~Ramsey}
\affiliation{Los Alamos National Laboratory, Los Alamos, New Mexico 87545, USA}

\author{A.~L.~Roberts}
\affiliation{Los Alamos National Laboratory, Los Alamos, New Mexico 87545, USA}

\author{R.~Schmid}
\affiliation{W.~K.~Kellogg Radiation Laboratory, California Institute
  of Technology, Pasadena, California 91125, USA}

\author{G.~M.~Seidel}
\affiliation{Department of Physics, Brown University, Providence,
  Rhode Island 02912, USA}

\author{D. Wagner}
\affiliation{Department of Physics and Astronomy, University of
  Kentucky, Lexington, Kentucky 40506, USA}

\author{W. Yao}
\affiliation{Oak Ridge National Laboratory, Oak Ridge, Tennessee
  37831, USA}

\date{\today}

\begin{abstract}
A new experiment to search for the permanent electric dipole moment of
the neutron is being developed for installation at the Spallation
Neutron Source at Oak Ridge National Laboratory. This experiment will
be performed in liquid helium at $\sim 0.4$~K and requires a large
electric field ($E\sim 75$~kV/cm) to be applied in liquid helium. We
have constructed a new HV test apparatus to study electric breakdown
in liquid helium. Initial results demonstrated that it is possible to
apply fields exceeding 100~kV/cm in a 1~cm gap between two
electropolished stainless steel electrodes 12~cm in diameter for a
wide range of pressures.
\end{abstract}

\maketitle

\section{Introduction}
\begin{figure}[tb]
\centering
\includegraphics[width=3.5in]{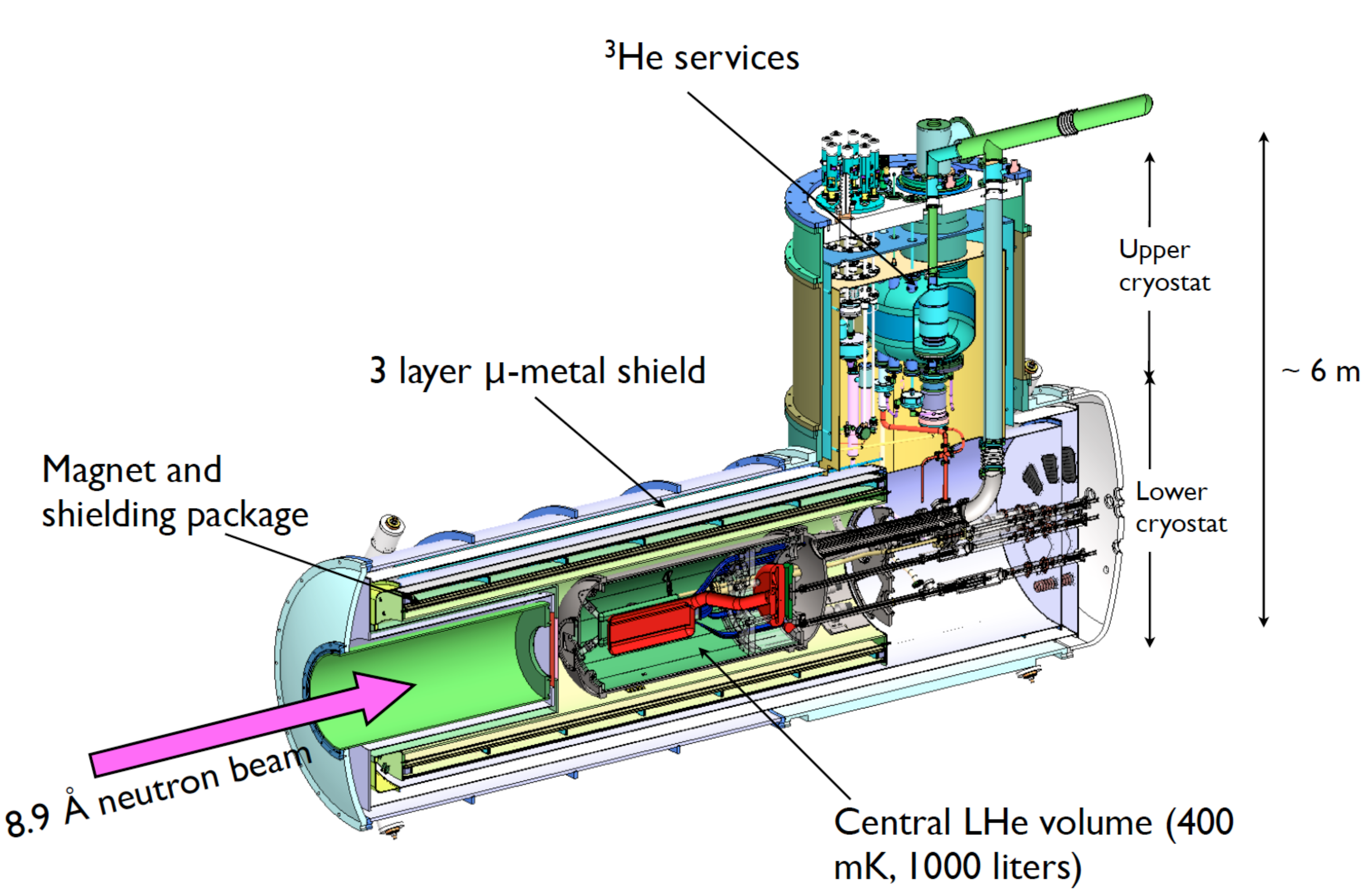}
\caption{A schematic of the apparatus for the SNS nEDM experiment, as
  is currently designed\label{fig:SNSnEDM}}
\end{figure}

A nonzero permanent electric dipole moment (EDM) of a nondegenerate
state of a system with spin $J \neq 0$ violates time reversal
invariance as well as invariance under parity operation. The time
reversal invariance violation implies a CP violation through the $CPT$
theorem. Given the smallness of the standard model CP violating
contributions induced by quark mixing, an EDM is a sensitive probe of
new physics. A new experiment to search for the permanent EDM of the
neutron, based on the method proposed by Golub and
Lamoreaux~\cite{GOL94}, is being developed to be mounted at the
Spallation Neutron Source at Oak Ridge National Laboratory, with a
sensitivity goal of $\sim 5\times 10^{-28}$~$e\cdot$cm, an improvement
of roughy two orders of magnitude over the current
limit~\cite{BAK06}. For more details of the current status of this SNS
nEDM experiment, see e.g. Ref.~\cite{ITO07}. A schematic of the
apparatus for the SNS nEDM experiment is shown in
Fig.~\ref{fig:SNSnEDM}.

\section{HV requirements}
The SNS nEDM experiment requires that a high, stable electric field
($\sim 75$~kV/cm) be applied in the region inside the two measurement
cells that are sandwiched between electrodes. The measurement cells
and electrodes are immersed in 0.4~K liquid helium (LHe). The
measurement cells, which store ultracold neutrons, are filled with
isotopically pure liquid $^4$He at $\sim 0.4$~K (the relative $^3$He
concentration $\sim 10^{-10}$). The measurement cells are made of PMMA
and are 10.16~cm$\times$ 12.70~cm $\times$ 42~cm in outer dimension
with a wall thickness of 1.2~cm. The electrodes, roughly
$10$~cm~$\times$~40~cm$\times$~80~cm in size, are made of PMMA coated with
a material that needs to meet various requirements related to
electrical resistivity, neutron activation properties, and magnetic
properties, etc. The leakage current along the cell walls need to be
minimized.


\section{Some general remarks on electrical breakdown in LHe}
Electrical breakdown in LHe, or more in general electrical breakdown
in any dielectric liquid, is rather poorly understood. Data exist on
electrical breakdown in LHe for temperatures of $1.2-4.2$~K (with the
bulk of data being taken at 4.2~K) mostly at the saturated vapor
pressure (SVP) for various electrode geometries, including
sphere-to-sphere, sphere-to-plane, and plane-to-plane. However, there
is little consistency among the data, and therefore there is no
consistent theoretical interpretation.

However, a rather simple consideration of the mean free path of ions
in LHe (electron bubbles and snow balls) and the electric field
strength necessary to accelerate them to an energy sufficiently high
to generate subsequent ionization leads to a conclusion that the
intrinsic dielectric strength of bulk LHe is greater than 10~MV/cm, a
field much higher than breakdown fields experimentally observed. This
leads to the following generally-accepted picture for the mechanism of
generation of electrical breakdown in LHe:
\begin{enumerate}
\item A vapor bubble is formed on the surface of the electrode,
  e.g. by field emission from roughness on the cathode
\item The vapor bubble grows, presumably by heating of the gas by
  accelerated electrons and evaporation of the liquid as a result, and
  forms a column of gas reaching from one electrode to the other
\item Electrical breakdown occurs through the gas column
\end{enumerate}
It follows that the parameters that can affect the breakdown field
strength include: (i) electrode material, in particular the surface
properties,and (ii) LHe temperature and pressure. In addition, because
electrical breakdown is a stochastic process, the size of the system
affects the breakdown field strength and its distribution. See
e.g. Ref.~\cite{WEB56}.

\section{R\&D approach}
The consideration given above indicates that the R\&D for the SNS nEDM
experiment requires a study of electrical breakdown in LHe in a
condition (i.e. temperature, pressure, size) as close as possible to
that expected for the SNS nEDM experiment, using suitable candidate
materials.

It is also very important to study the effect of the presence of a
dielectric insulator sandwiched between electrodes, as such will be
the geometry for the SNS nEDM experiment. Note that even in a room
temperature vacuum system, electric fields exceeding a few 100~kV/cm
are possible when there is no insulator between the two electrodes. 
In a study performed using a room temperature vacuum
aparatus~\cite{GOL86} similar to those used in the previous nEDM
experiments (such as Ref.~\cite{BAK06}), the electric field was
limited to $\sim 30$~kV/cm due to the presence of the UCN confining
wall that was sandwiched between the two
electrodes~\cite{GOL86}\footnote{In the actual nEDM experiment, the
  achievable field was further lowered due to other factors}.  Field
emission at the cathode-insulator junction is thought to be
responsible for initiating breakdown~\cite{KOF60}, which we expect to
be suppressed at cryogenic temperatures.

In order to study the relevant aspects of electrical breakdown in LHe
with a goal of establishing the feasibility of the SNS nEDM experiment
as well as guiding the design of the apparatus, we constructed an
apparatus called Medium Scale HV (MSHV) Test Apparatus, which is
described below. 

\section{Medium scale HV test apparatus}
\begin{figure}[tb]
\centering
\includegraphics[width=3.5in]{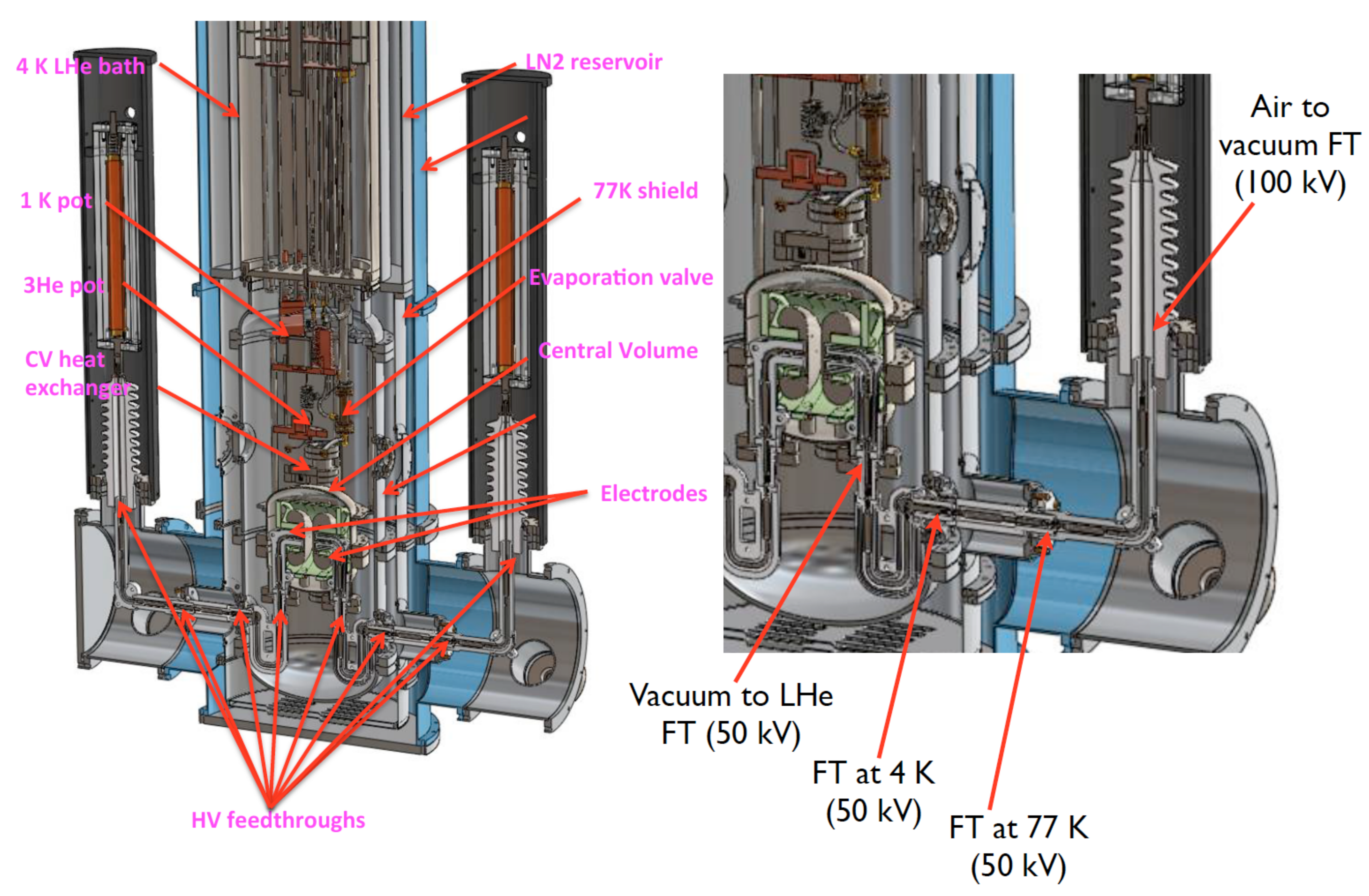}
\caption{A schematic of the MSHV system\label{fig:schematic}}
\end{figure}

The purpose of the MSHV system is to study electrical breakdown in LHe
in a condition approximating that of the SNS nEDM experiment using
suitable electrode candidate materials. The lowest operating
temperature of the MSHV system is designed to be 0.4~K, corresponding
to the operating temperature of the SNS nEDM experiment. Since we
expect that the pressure is an important parameter affecting the
breakdown field strength in LHe, the MSHV system is designed so that
the pressure of the LHe volume in which the electrodes are placed can
be varied between the SVP and 1~atm. Note that the SVP is $\sim
10^{-6}$~torr at 0.4~K. The size of the LHe volume was determined as a
compromise between the following two competing factors:
\begin{itemize}
\item Short turnaround time of the system ($\sim 2$~weeks) to allow
  multiple electrode material candidates to be tested.
\item Size large enough to give information relevant for the SNS nEDM experiment.
\end{itemize} 
The electrodes are 12~cm in diameter. The gap size is adjustable
between 1 and 2~cm. Each dimension is within a factor of 10 of the SNS
nEDM experiment's HV system. A schematic of the MSHV system is shown
in Fig.~\ref{fig:schematic}.

The Central Volume (CV), a 6-liter LHe volume that houses the electrodes, is
cooled by a $^3$He refrigerator. 

HV between $-50$~kV and $+50$~kV can be provided to each electrode
through a HV feed line. The HV feed lines are made of thin wall
stainless steel tubing and are thermally anchored at the LN2 heat shield
and at the 4~K heat shield, in order to minimize the heat leak to the
HV electrodes. Heat leak to the HV electrodes can cause vapor bubbles
to be created on the surface of the electrodes, which in turn can
initiate electrical breakdown, potentially leading to erronoues
results. 

Commercially available models are used for all the HV
feedthroughs. For the air-to-vacuum feedthroughs, CeramTec Model
6722-01-CF feedthroughs, rated for 100~kV, are used. For all other
feedthroughs, including ones on the CV that need to be superfluid
tight, CeramTec Model 21183-01-W, rated for 50~kV and for LHe
temperature operation, were chosen because the spatial limitations did
not allow larger sized feedthroughs. In an offline test of these 50~kV
feedthroughs, we found that, after proper cleaning, they can withstand
a HV up to $\sim 90$~kV with a leakage current of less than 1~nA in
vacuum when cooled to 77~K.

Most of HV components were tested in a separate system for holdoff
voltage before being installed into the MSHV system. Our experience is
that a component that functions in a room temperature vacuum generally
functions in LHe.

\begin{figure}[tb]
\centering
\includegraphics[width=3.5in]{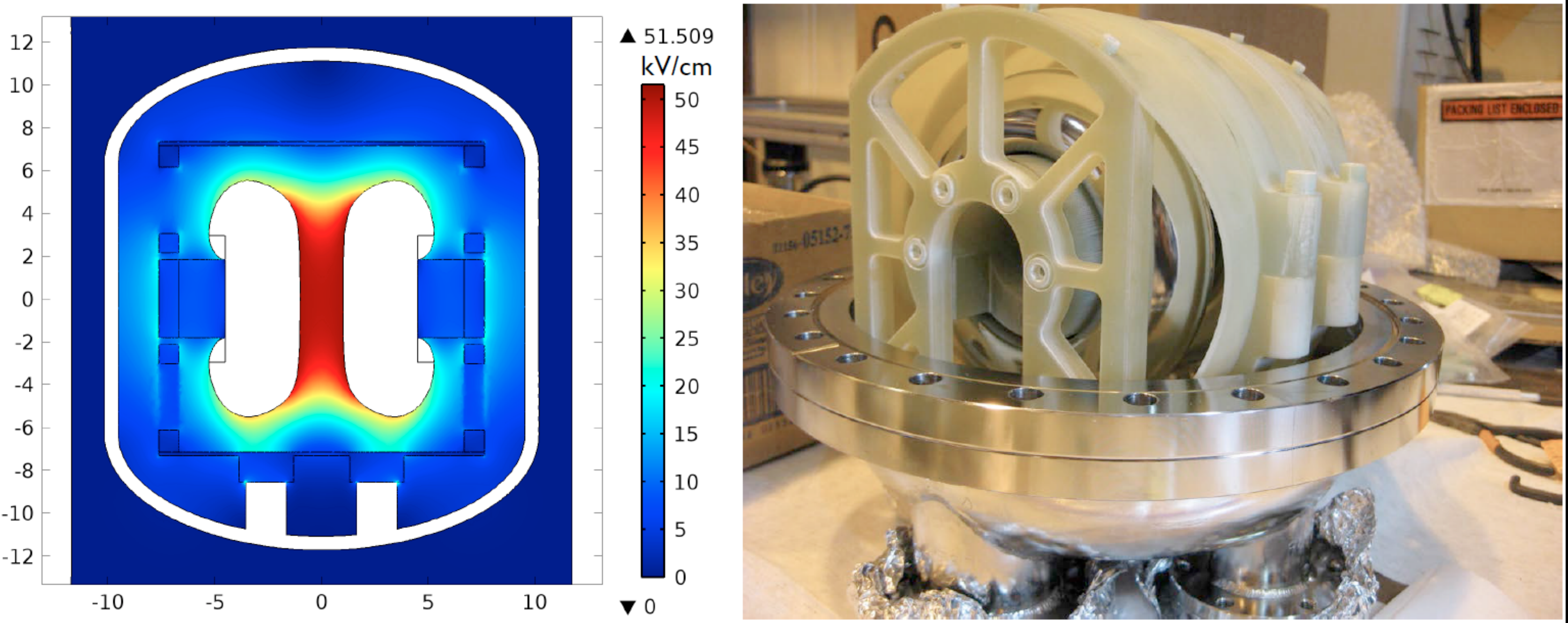}
\caption{Rogowski electrodes for the MSHV system. Left: FEM
  calculation of the electric field inside the CV showing a uniform
  field in the gap. Right: a photograph of the Rogowski electrodes
  installed in the CV. \label{fig:electrodes}}
\end{figure}

The initial electrodes are made of electropolished stainless steel and
have the so-called Rogowski profile~\cite{COB58}, which provides a
uniform electric field in the gap and ensures that the gap has the
highest field in the system (see Fig.~\ref{fig:electrodes}).

\section{Progress to date}
We have successfully constructed and commissioned the MSHV system. We
have demonstrated that the CV can be cooled to 0.4~K with it filled
with LHe.
We have also demonstrated that the
pressure inside the CV can be varied and controlled easily.

\begin{figure}[tb]
\centering
\includegraphics[width=3in]{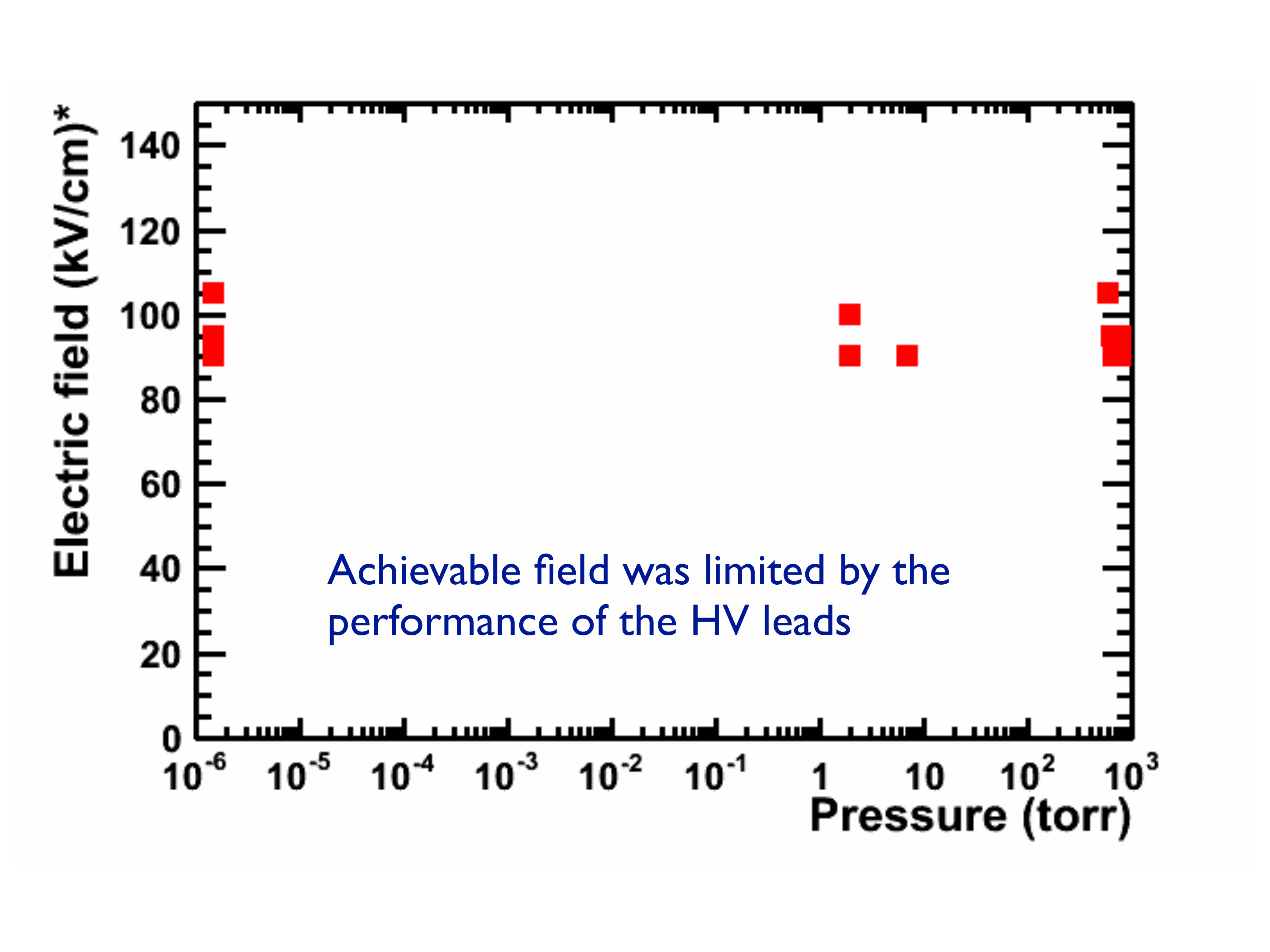}
\caption{Electric field strength achieved in a 1~cm gap between two
  electropolished stainless steel electrodes 12~cm in diameter. Note
  that the achievable field was limited by the performance of the HV
  leads. The hydrolic pressure of LHe is $\sim
  0.1$~torr/cm\label{fig:results}}
\end{figure}

In addition, we have demonstrated that an electric field exceeding
100~kV/cm can be stably applied in a 1-cm gap between the two 12-cm
diameter electrodes made of electropolished stainless steel for a wide
range of pressures (see Fig.~\ref{fig:results}). The achievable field
was limited by the performance of the HV leads. No breakdown was
observed in the gap between the two electrodes. Also the leakage
current between the two electrodes was measured to be less than 1~pA
at a 50~kV potential difference. This was measured with one of the HV
electrodes set to the ground potential to avoid the leakage current in
the HV cable and feedthroughs dominating the measurement.

\section{Summary}
For the HV R\&D for the SNS nEDM experiment, we have constructed a new
HV test apparatus to study electrical breakdown in LHe. Initial
results demonstrated that it is possible to apply fields exceeding
100~kV/cm in a 1~cm gap between two electropolished stainless steel
electrodes 12~cm in diameter for a wide range of pressures.

\end{document}